\documentstyle[editedvolume,numreferences]{crckapb}%
\def \Z {\rlap{\rm Z}\mkern3mu{\rm Z}}
\def \CV {{\cal V}}
\def \CF {{\cal F}}
\def \R {\rlap{\rm I}\mkern3mu{\rm R}}
\def \C {\rlap{\rm C}\mkern3mu{\rm I}}
\def \CM {{\cal M   }}
\def \CA {{\cal A   }}
\def\lsemidir{\mathbin{\hbox{\hskip2pt\vrule height 5.7pt depth -.3pt
width .25pt\hskip-2pt$\times$}}}

\def\sd{{\lsemidir}} 

\begin{opening}
\title{DUALITIES IN THE CLASSICAL SUPERGRAVITY LIMITS}
\subtitle{Dualisations, dualities and a d\' etour via 4k+2 dimensions}
\runningtitle{MORE DUALITY SYMMETRIES}
\author{B. L. Julia}
\institute{Laboratoire de Physique th\' eorique de 
l'Ecole Normale Sup\' erieure, CNRS \\
 24 rue Lhomond 75005 Paris FRANCE.}
\end{opening}
\begin{document}

\begin{abstract}
Duality symmetries of supergravity theories are powerful 
tools to restrict  the number of possible actions, to link different 
dimensions and number of supersymmetries and might help to control 
quantisation.  (Hodge-Dirac-)Dualisation of gauge 
potentials exchanges Noether and topological charges, equations of 
motion and Bianchi identities, internal rigid symmetries and gauge
symmetries, local transformations with nonlocal ones and most exciting 
particles and waves. 
We compare the actions of maximally dualised supergravities 
(ie with gauge potential forms 
of lowest possible degree) to the non-dualised actions coming from 11 (or 10)
dimensions by plain dimensional reduction as well as to other 
theories with partial 
dualisations. The effect on the rigid duality group is
a kind of contraction resulting from the elimination of the unfaithful 
generators associated to the (inversely) dualised scalar fields. 
New gauge symmetries are introduced  by these (un)dualisations and it is 
clear that a complete picture of duality (F(ull)-duality) should include all 
gauge symmetries 
at the same time as the rigid symmetries and the spacetime symmetries. 
We may read off  some properties of F-duality
on the internal rigid Dynkin diagram: field content, 
possible dualisations, increase of the rank according to the decrease
of space dimension... Some recent results are 
included to suggest the way towards unification via a universal
twisted self-duality (TS) structure. The analysis of this structure
had revealed several profound differences according to the parity mod 4 of the 
dimension of spacetime (to be contrasted with 
the (Bott) period 8 of spinor properties).
\footnote{ After the original lectures were delivered at this Carg\` ese school in May 1997, various developments 
have been presented at the  Neuch\^ atel Workshop ``Quantum aspects 
of Gauge theories, Supersymmetry and Unification" in September 1997 and at the
Trieste Conference on ``Superfivebranes and Physics in 5+1 dimensions" 
  in April 1998. Work supported in part by EEC under TMR contracts 
ERBFMRX-CT96-0012 and -45}
    
\end{abstract}

\section{Introduction}
The duality symmetries are invariances of equations of motion 
and even sometimes, for instance in odd dimensions, 
symmetries of a suitable action. The case of $4k+2$ dimensions seems to be 
related to classical   Lie-Poisson actions leading to Quantum groups, indeed
the latter were discovered in 2 dimensional theories.
The duality group gets its name from some of its elements that 
are actually Hodge dualities like in Dirac's famous analysis of the 
exchange of electricity and magnetism in four dimensions which permutes
strong and weak coupling expansions. This is also  the nature of so-called
discrete S-dualities in quantum four dimensional heterotic string theories as 
discussed 
by A. Sen. The $SL(2,\R)$ symmetry in type IIB theory in 10  dimensions is also
called S duality as it exchanges weak and strong string couplings. 
The string coupling is affected via the dilaton field. In four dimensions
these symmetries involve Hodge duality in the target spacetime.

 In string theory the so-called  T-dualities do
exchange strong and weak couplings of the sigma model on the worldsheet. 
They can be seen
as generalisations of the Kramers-Wannier symmetry of the Ising model 
that permutes inverse temperature (or equivalently the euclidean 
period of time) and its inverse. T-dualities may also be
realised as Hodge dualities but on the 2d worldsheet,
as was shown by Buscher when the 
target space admits isometries. A generalisation called Lie-Poisson T-duality
relaxes slightly the latter condition.

Three years ago Hull and Townsend unified the two main kinds of dualities
(S and T) inside a much larger if conjectural discrete
U-duality group of the quantum supergravity and string 
theories. These classical theories are now believed to be inequivalent limits
of a quantum model called M-theory after Witten's discovery 
of the correspondence between both formulations. The string coupling constant
becomes geometrical, essentially equal to the length of the compactification 
circle along the eleventh dimension. An important 
point is that all solitonic excitations should be included in 
the quantum theory, as well as their duals which include fundamental strings 
in ten dimensions or  fundamental membranes in 11. 

Now the proposed U-duality groups are discrete subgroups of the duality
symmetries of the equations of motion of maximally dualised supergravities 
which have been  known since about 1980.
We shall use the same letter for the discrete subgroup and the Lie group when
it is clear from the context which one it is. 
The first classical  supergravities under consideration 
were the fully dualised toroidal dimensional reductions of 
11 dimensional (ie type II or maximal or $N_4=8$) 
supergravity. We shall only briefly mention their
truncations to pure ($N_4 \le 6$) supergravities. 
Type I supergravities possess also 
interesting  duality symmetries of the classical equations of motion
 first studied by Chamseddine \cite{cham};
considered together with those of type II \cite{jchi}
they strongly suggest that the two simply laced hyperbolic Kac-Moody algebras
of maximal rank (equal to 10) should appear as symmetries of some huge space
covering the set of unidimensional classical solutions of supergravity 
theories. Although there are only indications for 
that yet, let us give a name to these hyperbolic Kac-Moody groups: $E_{10}$ and
$HD_{10}$ (called overextended $D_8$ in \cite{jchi}),
they  correspond to type II resp. type I.
It would be
interesting to
accommodate the heterotic theories in the hyperbolic game \cite{hm}. The
situation there is still moving with hyperbolic Kac-Moody algebras and
generalised Kac-Moody algebras (including Borcherds algebras) appearing in
toroidal compactifications.
 
What has been established and extensively studied is the
occurrence of infinite dimensional symmetry groups in the reduction to
two dimensions, there the affine Kac-Moody extension $G^{(1)}$ enlarges
$G$, the corresponding maximal U-duality group in three dimensions. 
In two dimensions scalar
potentials (fields) are dual to scalar fields, as a result the exchange 
between gauge and internal (rigid) symmetries does not take place; 
the duality group acting on the appropriate set of fields 
covering the set of solutions becomes infinite dimensional. 
It is also an important problem to describe precisely and uniformly,
namely for all dimensions, what the discrete (infinite) groups
of U duality are. They are most probably groups over 
$\Z $, the rational integers, as defined by Chevalley.
This conjecture is nicely compatible with the observation that any $\R$ 
factor group of the classical duality group has no infinite discrete duality 
analogue and disappears from the  U-duality group at the quantum level.

In section 2 we shall recall the ``silver rules" of supergravities and the
building blocks of the U-groups. The latter use three ingredients: the 
dilatonic 
rescaling symmetry one obtains in 10 dimensions, the ``Ehlers phenomenon" 
where a scaling symmetry becomes a whole $SL(2)$ and finally the fusing 
together
of such $SL(2)$'s with the expected symmetry $GL(D-d)$ that comes from the 
dimensional reduction from $D$ to $d$ dimensions into a simple rigid
internal U-duality group. Given $G$ the U-duality group, the three silver 
rules are that the scalar fields 
parameterise a symmetric space of the noncompact type $K\backslash G$ where 
K is the maximal compact subgroup of G, that equations simplify dramatically
upon restoring a local gauge invariance under K and finally that in even 
dimensions $d=2f$ the equations of motion of the field strengths of order
$f$ and their Bianchi identities are unified in a twisted self-duality 
equation.
$$* S.\CV.\CF =  \CV.\CF$$
where $S$ is an invariant operator acting on the appropriate representation
of $K$ and where 
$\CV$ is the coset representative of the scalar field transforming under $G$
on the right contragrediently to $\CF$  on the left, and  transforming
also under the local gauge group
$K$  on the left. The case of timelike compactification is somewhat  
different \cite{hull} but it is important for Euclidean signature,
see \cite{hagi}. The maximal compact subgroup is 
replaced in general  by a noncompact subgroup and the quotient has a  rather 
problematic topology.  

In section 3 we proceed to study the effects  of dualisations beginning with a
comparison between pure gravity reduced from 4 to 3 dimensions and its 
dualised theory with Ehlers symmetry acting in a local way as a rigid 
$SL(2,\R)$ invariance of the action. In the theory of
integrable Hamiltonian systems, Baecklund transformations (for instance Miura 
transformations) may exchange 
solutions of one system  with those of another. Here dualisations are
discrete duality transformations, essentially Legendre transformations, 
that modify the (perturbative) 
field contents of the action and the analogy is useful. 
Then we analyse in detail the case of maximal supergravity, also in three
dimensions. It may seem at first that as soon as one leaves the simple case of
maximally dualised theories one is  in danger of loosing oneself among all the 
possibilities. It is not so, the  key choice is to pick a grading of the root 
 space, for instance along a particular root, one can then show how 
undualisation of some scalar
fields belonging to the corresponding 
highest level (or levels) do actually reduce the
U-symmetry by changing the dimension of (and simultaneously contracting ie 
partially abelianising) the 
remaining subalgebra.
In passing we note that large abelian subalgebras of dualities do occur
and in fact those of maximal dimension as classified by Malcev \cite{mal} for
the case of complex Lie algebras (or their normal real forms) do appear.
In the case at hand there can be 36 commuting generators in the maximally
noncompact (so-called split
or normal) real form of $E_8$: $E_8(+8)$. This is to be contrasted with the compact
form situation where the maximal abelian dimension is the rank.

In the fourth section we shall try to learn about the higher form fields
from the Dynkin diagram of the U-group. It turns out that one can read off 
from the diagram the number of forms of various degrees, because they belong 
to very specific fundamental representations of $G$. In the case of maximal 
supergravities the Dynkin diagram reflects the possibility of 
dualisation between higher form fields by the existence of some outer 
automorphism. In general and most 
importantly it suggests that one should unify internal rigid symmetries with  
diffeomorphisms (or at least the $GL(D-d)$ subgroup)
in a larger group of rank at least eleven (or twelve?).  
This can be applied to type IIA, IIB or I. In fact we find a purely group 
theoretical version of the Horava Witten orbifolding relating type II and 
type I string theories, it corresponds to using a Cartan involution of the 
U-duality in any dimension. The heterotic duality groups are 
interesting too, they are non-split real forms in which one still  
recognizes the expected linear diffeomorphism symmetries, but  they will be 
treated separately. Note that the non-maximal
supergravities also lead to non-split real forms of the duality groups 
\cite{julia}.

Then we shall present the general analysis of duality symmetries in curved space
with either Lorentzian or Euclidean signature. The matter will be taken in an 
N-plet of middle rank ($f=d/2$)-forms plus sufficiently many  scalar fields. 
Since our paper \cite{CJLP1} appeared 
on the hep-th archive we realised that the general Lorentzian case seems to 
have been investigated by Tanii \cite{Ta}, some partial results on 
the 2 dimensional case were also obtained in \cite{CFG}. The 4 dimensional case
was systematically analysed in \cite{GZ} but as one might expect the $4k+2$
dimensional case is quite different. The fact that self-duality becomes 
possible is not directly relevant here. We have also
constructed constrained actions that preserve the U-symmetry and allow to 
simplify the computations by doubling the set of fields (third silver rule).
  
In the last section I shall summarize two research projects I have been 
working on for 
the past few years. On the one hand I want to stress the importance of the 
existence of ``complementary " classical limits appropriate to different 
experimental situations. This may lead to  some clarification of the 
formulation of Quantum Mechanics. There is no classical world only classical 
approximations: the limit of the Planck constant tending to zero is to be 
defined by a dimensionless criterion but more importantly by specifying what 
is being kept fixed. Particle limit and wave or classical field limit are in 
{\it duality}.   
On the other hand I would like to point out more publicly than before
the intimate relation between intersection theory, fermionisation 
and charge quantization of dyons. The fundamental differences among even 
dimensions between $d=4k$ and $d=4k+2$ will be discussed. The idea is that 
well defined statistics
is what saves locality in theories of extended objects, it is associated to 
a charge quantization condition in two dimensions. But the latter has to be 
symmetric to 
allow for chiral (self-dual) particles \cite{SW} \cite{BMT}. We show that 
higher dimensional fermionisation requires a d\' etour through 6 (or 10)
dimensions or twisting in 4d by some internal symmetry
in the sense of section 2. 
In  guise of conclusion I give a preview of a forthcoming paper
where it is shown that indeed the twisted self-duality equation is universal
excluding for the time being the graviton. There is a candidate for F-duality 
that brings us into the realm of graded  superalgebras. 
\bigskip 

 \section{Silver rules}

In the quest for supergravity actions the use of extra-dimensions of
space made the deformation or so-called Noether method significantly 
simpler by reducing the number of scalar fields to zero in the best cases
and thus avoiding nonpolynomial expressions. The idea of higher dimensions 
is in fact quite natural in the context of extended Poincar\' e superalgebras,
in particular the doubling of the number of fermionic charges can result from
the addition of 2 dimensions. Interest in four-dimensional supergravity with
the maximal possible number of supercharges (32) led us to consider ten 
dimensions. In
fact we went directly to eleven dimensions with Majorana spinors as 
suggested by the 
spectrum analysis by Nahm at the linear level but also by practical 
considerations. 
No scalar fields are left in eleven dimensions and the spectrum is 
amazingly simple: a graviton, a gauge three-form and the Rarita-Schwinger field.
Furthermore dimensional reduction is easier than its converse group 
disintegration \cite{julia}. For instance the theory with $24$ supercharges 
in 6 dimensions predicted in \cite{julia} was only constructed 16 years later
\cite{FK}. Returning to eleven dimensions, the requirement of local 
supersymmetry
leads to a unique action for this set of classical fields. After toroidal 
compactification one obtains
the expected $N=8$ supergravity in 4 dimensions but also peculiar duality 
symmetries. Indeed in dimension $d$ one discovered  (maximally) noncompact symmetry 
groups $E_{11-d}(11-d)$ of the toroidally reduced theories;
the number between parentheses is the real rank to be defined 
in the next paragraph. 

We found ourselves in a situation similar to that of general relativists in 
front of the Ehlers $SL(2,\R)$ symmetry or its extension for stationary electrovacs 
$SU(2,1)$. In the 
absence of conceptual understanding we began a systematic analysis of the symmetry
Lie algebras for all dimensions and all number of supersymmetries. This led to a 
list of noncompact symmetric spaces, ie cosets of the form $K\backslash G$
of the U-symmetry groups $G$ by their maximal compact subgroups $K$. 
The real rank of the coset is the maximal dimension of the subspace of a 
Cartan subalgebra orthogonal to the compact directions (for short noncompact Cartan 
generators). The real rank $r$ is equal to the rank $l$ 
for a normal (also called split) real form, this is the case for
the descendants of type I or II supergravities in 10 dimensions. It turns out that 
in all cases
the real rank increases by one upon each step of dimensional reduction. One can 
also check that
$d+l-N'_4 =4$ is constant in the disintegration triangle \cite{julia} 
for type II(A) and other pure (in four dimensions) 
supergravities in various dimensions; $N'_4$ is the number of 
supersymmetries in four dimensions except for the maximal case where it is 
equal to seven. For Chamseddine's type I 
disintegration column, the constant is equal to 7 instead of 4.  
Now the scalar fields parameterize this coset and other fields transform as representations of
$K$ or if one prefers of $ G$ through the scalar group element which intertwines
between $K$ and $G$ (we do not 
consider spinor fields here). It is quite 
suggestive to think of (the scalars as) a moving frame exchanging the ($K$)
tangent space-Lorentz indices with ($G$) world indices.

The split or maximally noncompact 
real form of a complex simple Lie group 
is the generalisation to a general Lie group of  $SL(2,\R)$ inside
$SL(2,\C \,)$. One 
simply defines the split form as generated by a number of copies of $SL(2,\R)$ 
equal to the rank $l$ and a finite set of relations, the Serre relations, which restrict their
simple and multiple commutators. The same relations hold over the real numbers for the 
split form as over the complex for the complex form. Another characteristic of
the split form is that it admits a
Cartan subalgebra of $l=r$ commuting, noncompact but ad-diagonalisable generators. The rest of 
the generators come in real pairs associated to opposite roots and
closing on  Cartan generators to form again  copies of $SL(2,\R)$. 
In the non-split case, the rank $l$ is strictly larger 
than the number $r$ of linearly independent noncompact Cartan generators.
However if we relax the diagonalisability assumption the number $a$ of linearly independent 
commuting generators might be strictly larger than $r$ and even than $l$.
We shall see examples of this in the next section for real split forms. The split real form is
also characterized by having the smallest compact subalgebra and consequently the largest 
dimension for the coset space $K\backslash G$. 

\bigskip

This symmetric space structure $K\backslash G$ for the moduli space of
scalar fields in the maximally dualised form of pure supergravity theories 
constitutes the first silver rule. Some of the gauge forms descending
either from the gauge three form or from the metric, namely the Kaluza-Klein one 
forms, can
be simultaneously dualised when their potentials can be covered with derivatives
despite the presence of the Chern-Simons-like term in 11 dimensions. In fact all those potentials that
are of degree $d-2$ can be dualised into scalars in the case of pure 
supergravities, this we shall  call maximal dualisation
in the scalar sector. That particular form of the action has the maximal rigid
internal symmetry.

The second silver rule is the generalisation of the orthonormal moving frame technique to internal
symmetries. In General Relativity the metric is given on a manifold, it does not depend
on any frame choice but its components depend on coordinate choices in a tensorial way. 
Alternatively
the same local information can be obtained from a frame of orthonormal 1-forms, but they are only defined up to 
a gauge Lorentz transformation. In our scalar manifold case,  $K\backslash G$ 
can be parameterised either
as a manifold by ignoring the coset structure or alternatively as the base of 
a principal bundle
the group $G$ itself. Then the subgroup action (on the left here) is a gauge symmetry that
compensates for the extra freedom that was added to make the $G$ symmetry manifest. 
Typically, our algebraic power being limited, the coset representatives we 
 encounter first are 
very often in the triangular or Borel and more generally in the solvable $K$-gauge
\cite{T}. One reason is 
that in that gauge the exponential parameterisation is bijective but more 
important probably is the fact 
that unipotent elements and their inverse are parameterised polynomially. 
The search for hidden 
symmetries is consequently the restoration of the gauge freedom or at least, and this is easier,
the enlargment of the Borel symmetry to the full $G$ invariance. For instance 
in two dimensions only the second process has been realised yet.

The third silver rule of supergravity is that the middle degree field strengths are self-dual; 
more precisely
their equations of motion and their Bianchi identities can be unified in a covariant set 
of Bianchi identities for a doubled set of fields restricted by a twisted self-duality condition 
so that the original $n$ second order equations have been encoded into a set 
of $2n$ first order 
ones. This was first discovered in 4 dimensions with Minkowskian signature such that the
square of the Hodge dual is equal to minus one: clearly the self-duality needs 
a twist in the form of a $G$ invariant operator $S$
acting on the representation $2n$ and of square equal to
minus one in four dimensions in order to compensate for the previous minus 
sign. As an illustration let us recall that maximal supergravity in 4 
dimensions has 28 vectors, with their dual potentials this makes 56. But this 
is precisely the dimension of the fundamental representation of $E_7$, it is 
symplectic and can be extended to a representation of $SP(56,\R)$.
 
It is an easy exercise to check that the square of the Hodge 
operator is equal to $(-1)^{(s-t)/2}$. Note that the difference $s-t$ also appears in the 
classification of Majorana spinors. In fact instantons require Euclidean 
(or (2,2)) signature in four
dimensions. But supersymmetry can provide some internal degrees of freedom to compensate for
the otherwise devastating minus sign, in other words it recalls its higher (10 
or 6) dimensional 
origin to allow for generalised or we shall say twisted self-duality (TS). 
Twist can  also be invoked to 
allow for generalised spin structure
or to define   a generalised Majorana condition.
This self-duality is a feature of toroidal compactifications as the appearance of bare potentials
rather than field strengths associated to nonabelian gauge theories or simply (charged) 
matter couplings seems to ruin the possibility of dualisation. 

We should be slightly more careful 
though as we know examples of dual pairs of theories one of which has bare potentials. For instance
the Freedman-Townsend-Thierry-Mieg dual sigma models are theories of two forms in four dimensions
with bare potentials. This is compatible with dualisation and even twisted self-duality
\cite{CJLP1}, \cite{CJLP2}. 
Indeed, in the most favorable case,  dualisation can use either one of two 
routes: one either introduces 
a Lagrange multiplier for the field strength Bianchi identity which becomes the dual potential,
this is possible provided no potential appears in the Lagrangian or in the 
Bianchi identity,
or one may use another first order formalism where now both the original potentials and field 
strengths are considered as independent variables. In the sigma model case the first route is 
taken when going from the scalar field description 
towards the two-form version but the reverse route is of the second type. This
reciprocity of the two routes occurs in general. In the dual sigma model 
case the alternative paths are not allowed. 
Furthermore doubling does not mean one can actually find a dual Lagrangian as 
the  example
of 11 dimensional supergravity shows \cite{CJLP2} and \cite{BBS}. There exists a 
doubled TS (twisted self-dual) formalism
but the 3-form cannot be integrated out to give the dual 6-form theory. 
 
 Let us now return to the general discussion of the third silver
rule of supergravity.
The structure is a mixture of $K$ and $G$ group theory. 
The potentials form a $G$-multiplet
but the field strength combinations that are self-dual are the $K$-multiplet. 
The key equation was 
already mentioned, it is $*S.\CV.\CF =  \CV.\CF$. This is to be supplemented by 
the $G$ tensor equation $\CF=d\CA$. This coexistence of differentiation in
curved space ($G$ representation) and self-duality in flat  or tangent space
($K$ representation) seems to be quite general, it holds for instance also 
 for various (super)brane actions and for Born-Infeld theories.

There are important remarks to make at this point. Firstly $S$ is 
an operator from the $K$ representation to itself. 
Secondly one should point out that
there have been two ways to write this self duality equation, we just gave the 
second one, the first one was given in \cite{CJ} and references therein: one
constructs the analog of the metric (the monodromy in integrable systems). In
the simple case of $SO\backslash SL$  we may define using a 
$SO$-invariant positive definite metric   
$$\CM_{MN}:=\CV_M^{t\, A} \eta_{AB} \CV^B_{\, N}.$$
 Then the TS equation reads
$$*\Omega^{PM}   \CM _{MN} \CF ^N = \CF^P.$$
The invariant tensor $\Omega$ is an invariant of the noncompact group $G$, 
for instance the symplectic form of $E_7$ in its 56 dimensional representation 
in the case of four spacetime dimensions.

How can we relate this formula to the previous one using $K$-tensors? 
It does not seem to have been written up in general in the litterature but we 
need to establish
$$ \CV_{\, M}^A \Omega^{MN} \CV^{tB}_N = S^A_{\, C} \eta ^{-1\, CB}:=\omega^{AB}$$
or alternatively
$$ \CV^{t\, -1} = \omega^{-1}.\CV . \Omega $$
choosing the identity matrix for $\CV$  we must identify $\omega$ and $\Omega$.
Things becomes even simpler if we use the operator $S$ and rewrite the above 
formula as $$\sigma(\CV) = S^{-1} \CV S $$
where $\sigma$ is the Cartan involution\footnote{Let us recall
that the maximal compact subgroup of a real simple Lie group can be defined
as the fixed point set of an involution
$\sigma$,  the Cartan involution of $G$. In the simple case under consideration
$\sigma(\CV) :=  \eta^{-1} . \CV^{t\, -1} . \eta$.}
The Cartan involution is  ``inner" in this representation, the condition for 
involutivity (that $S^2$ be central) is automatic here. 
This property of $\sigma$ has interesting 
consequences to which we hope to return in the future.
For a   general group, following \cite{BM}, we must introduce
$\sigma(\CV ^{-1})\CV $ to replace $\eta^{-1} \CM$. 

There is no one-to-one relation between the symmetry  property 
of  the twist-form $\omega$ and  
the parity of $(s-t)/2$. The latter is related to reality properties whereas
the former is a linear question. 
The symmetry of the twist can be found by inspection,
surprisingly it does not depend on the signature of spacetime. One instance of 
this ``signature blindness", 
namely the fact that duality transformations are canonical in $4k$ 
dimensions and not in $4k+2$ dimensions in Euclidean signature as well as
in the Minkowskian one, was 
actually discovered some time ago\footnote{in a discussion with S. Deser in 
Sept. 1996.}. Let us recall also that the fact that
 dualities are not canonical in 2 dimensions was realised before Quantum groups
were invented see for instance \cite{mac}.
In $4k$ Lorentzian dimensions one has no real ``self-dual" tensor but there is 
a Euclidean one and twist is 
antisymmetric, in $4k+2$ dimensions on the other hand twist is symmetric. The 
existence of a symmetric resp. antisymmetric bilinear, invariant 
form for an irreducible complex representation of a group that is equivalent to 
its contragredient (here by conjugation by $\omega$) makes it a ``real" resp. 
``quaternionic" representation \cite{Ti}. 

The analysis is complicated here by the fact that as we shall see the subgroup 
$H$ that replaces $K$ in the Euclidean case is non-compact. The properties of 
the twist $\omega$, the 
invariant metric $\eta$ and the operator $S$ are intertwined in the relation
$S:=\omega . \eta$. 
The origin of the two different symmetries of $\omega$ has not been completely 
clarified yet, 
but the detailed study of supergravities is quite suggestive. We shall review 
the systematic analysis of \cite{CJLP1} in section 4. 
For comparison, let us recall that half-spinor representations in even 
Lorentzian
dimensions are  complex if $d=4k$, real if $d=8k+2$ and quaternionic if 
$d=8k+6$. The distinction between quaternionic and real spinors is irrelevant
for quadratic expressions like the Ramond-Ramond bosons under consideration 
here. More relevant is the fact that the only symplectic fundamental 
representation of $SL(D,\C)$ is the self-dual tensor for $D=4k+2$ \cite{Ti}. 
Let us also note that the 
fundamental representations of $SP(2N,\C)$ and of $SO(2N,\C)$ are obviously
respectively symplectic and real.

Let us now review how U-dualities grow upon toroidal dimensional reduction.
In eleven dimensions the classical equations of motion admit an engineering
 scaling symmetry
because the dimensionful coupling constant appears as an overall factor in the 
action. In 10 dimensions this leads to an $\R$ internal and rigid symmetry.
Technically one must specify an (Einstein) frame. This noncompact symmetry 
rescales the dilaton or the length of the compactification circle. Beyond this 
one generator there is an automatic $SL(11-d)$ internal symmetry directly 
originating from the diffeomorphism symmetries of our starting point. The 
second fact to notice is that in 8 dimensions the scaling group becomes an
$SL(2,\R)$ factor. This is also the dimension in which the 4-form field strength
can be self-dual. More surprising still is the fusion into a simple Lie group 
of both ingredients, $SL(2,\R)$ and $SL(11-d)$  below 8 dimensions. 

This discussion can 
be adapted to the type I family. The starting point is in 10 dimensions 
\cite{cham} with one dilatonic symmetry. Now in 8 dimensions a second 
$SL(2,\R)$ factor appears beyond the expected $SL(10-8, \R)$, it incorporates 
one of the dilatonic symmetries of the 9 dimensional model. 
In lower dimensions the duality symmetries are the 
now familiar $SO(10-d,10-d)\times \R$ enlarging $SL(10-d,\R)$.  In 4 dimensions 
the last dilatonic 
subgroup becomes a full $SL(2,\R)$ factor by dualisation of the two-form
potential. 
Finally in three dimensions the group becomes simple namely $SO(8,8)$.

Until now we focussed our attention on  the maximally dualised theories. 
But if instead of deciding to lower as much as possible the degrees of the forms
in the action by dualising, when possible, any k-form field strength to the 
lower degree
$(k'=d-k)$-form we decide not to do it or  to undualise some of the latter,
the internal symmetry group action on the scalar fields  that are being 
undualised is transmuted into a gauge group for the dual 
forms, at least above 2 dimensions. For the two-dimensional case see for 
instance \cite{BJ} and references therein. The purpose of the next section is 
to discuss more systematically 
these choices (to dualise or not to dualise). 

         \bigskip 
\section{Dualisations}

It may be advisable to begin with the simplest typical example rather than a 
list of definitions and formal properties. So let us follow J. Ehlers who 
recognized a $U(1)$ duality symmetry of Einstein's vacuum solutions in 4 
dimensions with one non-null Killing vector. One can view the situation as 
a fibration of spacetime over a three dimensional set of orbits. One question 
is whether the base inherits a geometrical structure namely a metric from the
original Minkowskian manifold. Locally the answer is yes, it is the beautiful
work of Kaluza (who actually developed the idea in 5 dimensions), it was
 completed by O. Klein (after Quantum Mechanics), Jordan, Thiry ...
In fact one may see that the Killing orbits being non-null the distribution of 
normal hyperplanes is transverse to the fibration, equivariant under the 
isometries by construction and thus defines a principal abelian connection  
(locally in all directions). Furthermore the length of the Killing vector 
defines a scalar field $C$ on the base. The construction of the geometry on the base
is given in \cite{RG}. The case of null Killing vectors is much more subtle see
\cite{JN} and references therein. 

Now as we started from 4 dimensional spacetime the two degrees of freedom 
of the 
graviton have become a scalar field plus  one polarisation of a vector gauge 
field (the connection we just presented). What are the symmetries? From 4d 
diffeomorphism invariance we expect a remaining scaling invariance of the 
cyclic (ie now internal) direction and of course the 
Maxwell-Weyl gauge invariance of the 
connection.  However it turns out that the connection 1-form can be dualised, 
in other words it can be exchanged for a dual scalar field $B$ 
by a Legendre transform. Then the miracle in 
today's state of affairs is that the two scalar fields do form a couple of 
coordinates for the Poincar\' e upper-half-plane. In other words they 
parameterise 
the symmetric Riemannian non-compact space $SO(2) \backslash SL(2,\R)$.  
Ehlers considered the $SO(2)$ subgroup of $SL(2,\R)$  but the full group is 
now  
called Ehlers' group. The action is pointwise as in a sigma model and rigid ie. 
independent on the position. But that means that it is actually non-local in 
terms of the original scalar $C$ and the original connection. Furthermore
the other two generators are the expected scaling and a new shift
of the second scalar $B$  which is defined up to an arbitrary additive
 constant by the Baecklund type formula:
$$ dA \propto *dB \, \, C^4 .$$ 
See for instance \cite{julia} for complete formulas. It is also established 
there that the Ehlers rotations act at the linearised level as helicity 
rotations.

To summarise, the maximal dualisation leads to  an unexpected symmetry that is 
nonlocal in terms of the original (non-dualised) fields. The undualisation of 
the new scalar field to the connection one-form hides the large symmetry but 
restores the gauge invariance of the connection field.

Our next example will be three dimensional maximal supergravity 
(type II but for our purpose
the fermion fields may be set to zero). The maximally dualised theory is in a 
topological background gravity and its dynamics is that of a Riemannian 
symmetric space sigma model of the non-compact type. The 128 bosonic degrees of
freedom span $SO(16)\backslash E_8$ or equivalently, but after gauge fixing the 
$SO(16)$ gauge invariance, they span
a Borel (or upper triangular) subgroup. In a 
maximally noncompact real form the Cartan generators may be chosen noncompact 
and together with the positive roots  they generate a solvable group that is 
univocally parameterised by the exponential map. The compact generators are the 
differences of positive root generators and their opposite and disappear 
in the gauge fixed description. Let us briefly recall the origin of the various 
scalar fields. The three form in eleven dimension reduces to 8x7x6/3!= 56 
scalars plus 28 vectors (to be dualised to scalars) and some non propagating 
components. The metric leads to 11-3=8 vectors (also dualisable) as well as 36 
scalars.  It turns out that these fields fit into the graded decomposition of 
the Borel subalgebra along the simple positive root labeled 10  
of the Dynkin diagram of $E_8$ in Figure 1. 

\centerline{
\begin{tabular}{ccccccccccccccccccc}\\
 $9$& &$8$& &$7$& &$6$& &$5$& &$4$& &$3$& &$(2)$ \\
 o&---&o&---&o&---&o&---&o&---&o&---&o&...&(o)\\
 &   & &   &$|$&   & &   & &   & &   & &  &\\
 &   & &   &o&   & &   & &   & &   & &  &\\
 &   & &   &$10$& & &   & &   & &   & &  &\\
\end{tabular}}
\bigskip
\centerline{Figure 1: type IIA symmetry, Dynkin diagram of $E_8$ with affine extension.}
\centerline{The label indicates the dimension at which each vertex appears. }
\bigskip

Let us recall that in the decomposition 
of a positive root into a linear combination of simple positive roots with 
non-negative coefficients there is a $\Z^8$ grading at our disposal. 
We claim that the simple root corresponding to the vertex 10 must be selected 
to find the symmetry of the bare reductions from 11 dimensional 
supergravity doing no dualisation at all. Firstly one must select the grading 
along that root, and then erase its vertex.
Namely there are  28  positive roots  of level 0 along that simple root
(plus 8 Cartan generators) all inside the Borel subalgebra of 
$gl(7,\R)$, as well as 
56 level 1 roots, 28 at level 2 and 8 at level 3. Clearly the level 2 and 3 root
generators form an abelian ideal because the level runs only up to 3. This means
in practical terms that the corresponding 36 scalars can be undualised to vectors. 
They actually originated as such from dimensional reduction, and $GL(7,\R)$ is the
almost obvious symmetry after strict toroidal dimensional reduction. 

We just gave the simplest example of the phenomenon: the big Borel duality
group of the 
maximally dualised theory is quotiented by the abelian ideal of the undualised
scalar generators to become typically a semidirect product, here (after 
restoring the full linear group) 
$GL(7,\R)\sd \R^{56}$. What has happened is that the 36 arbitrary constants  in 
the definition of the new scalar fields obtained by dualisations are not 
available before 
dualisations and their shifts by corresponding group elements drop out because 
the realisation  of the Borel group becomes non faithful; furthermore non 
commuting generators of $E_8$, like 
those at level 1, become commuting after undualisation because their 
commutators vanish by the previous mechanism. This phenomenon has been 
uncovered in 
\cite{CJLP1} but abelian ideals have many other uses see \cite{T}. 
If one wants to restore gauge invariance under the compact subgroup (second 
silver rule) one may 
alternatively describe the division by the ideal of the Borel group as a double 
coset construction for the full U-duality group \cite{CJLP1}.

Such undualisation along the root 10 of all the $E_{11-d}$ Borel subgroups
leads to symmetries of a similar type: semidirect products of the corresponding 
linear subgroup by some
abelian group which are the symmetries obtained from bare dimensional 
reduction from 11 dimensions. We did not give the details here but the
$SL(11-d)$ is visible as a  subdiagram of the $E_8$ Dynkin diagram that grows 
together with $E_{11-d}$ from the vertex labeled 9 towards the vertex labeled 
3  ...
For all dimensions one uses abelian ideals, those of  the maximal 
possible dimension  occur in various places as one can check following 
\cite{mal} who found them by inspection. A 
more conceptual discussion will be attempted in a further paper. But let us 
illustrate  their usefulness on more examples. 

Let us now turn briefly to three other natural choices of undualisations. We 
shall consider undualisation of all NS-NS fields (the even forms), 
it corresponds to exchanging the role of the simple 
roots 10 and 9  in Figure 1, namely to undualising generators whose root 
has the highest coefficient along the simple root 9. 
We could also consider type IIB supergravity in 10 
dimensions, reduce to 9 dimensions and below without dualising the 5 form 
field-strength. This series is obtained by using root 8 (and simultaneously 
root 
10) in a similar way as root 10 or 9 in the previous examples. The manifest and 
expected symmetry is now only $GL(9-d,\R)$. Finally the truncation 
of $E_{11-d}$ to $E_{10-d}$ can also be obtained by deleting the vertex labelled
$d$  at the end of the $SL(11-d)$ line inside ($E_{11-d}$) that shortens  when 
$d$ increases.
The origin of this property is that any vector in dimension $d+1$ gives 
rise to a new  scalar in dimension $d$. Malcev has shown that the maximal
dimension of a maximal abelian subalgebra of $E_7(7)$ is equal to  27, 
but 27=63-36 which corresponds precisely to the contraction of  $E_7(7)$ 
towards $E_6$. The rest of the $E$ family gives similar results.

The last series we would like to discuss is the type I family between 10 and 3 
dimensions. The diagrams are subdiagrams of Figure 2:

\centerline{
\begin{tabular}{ccccccccccccccccccc}\\
 $8$& &$7$& &$6$& &$5$& &$4$& &$3$& &$(2)$ \\
 o&---&o&---&o&---&o&---&o&---&o&...&(o)\\
 &   &$|$&   & &   & &   & &   &$|$&  &\\
 &   &o&   & &   & &   & &   &o&  &  \\
 &   &$9$ &   & & & &   & &   &$10$&  &  \\
\end{tabular}}
\bigskip

\centerline{Figure 2: type I symmetry, Dynkin diagram of $D_8$ with affine extension.}
\centerline{ }
\bigskip
Again the labels denote the dimension at which the Cartan generators appear,
sometimes with a full $SL(2,\R)$ but not immediately. 
We may conclude this section with the remark that there is
also a couple of undualisations that are easy to describe in the type I case,
namely those of the 2-form. They are done by the same type of contraction
as for type II by using the simple root labelled 10 in Figure 2.

\bigskip
\section{Higher order potentials}

In the previous section we have seen how specific locations on the Dynkin 
diagram of internal dualities correspond to specific sets of dualisations
and in the process of studying these we have seen that root 
labelled $d$ is associated to vectors. We shall now discover, in type II 
supergravities, that root 10 is associated to the 3-form, root 8 to the 
4-form and root 9 to the 2-form potentials. Let us consider for 
instance dimension 7, there are 4 vectors from the metric plus 6 from the 
3-form, together they build the representation 10 of $SL(5,\R)$ which 
corresponds to the fundamental dominant weight on vertex $d=7$. 
Similarly in dimension 6 the
3-form can (and should) be dualised to a 16th vector to implement the symmetry
$SO(5,5)$ and again this is the fundamental half spinorial representation 
associated to vertex $d=6$.  If the 3-form is associated to vertex 10 we might 
guess that the symmetry of the Dynkin diagram in 6 dimensions is responsible 
for the possibility of dualising it away.  

In dimension 2 one knows that the U-symmetry of type I supergravity is 
the affine $D_8^{(1)}$. Note that there is a Weyl reflection 
that exchanges the  highest root (it is the opposite of the root 
labelled (2) corresponding to the affine extension) 
with the simple positive root labelled 3. Roots 3 and (2) are 
connected by a simple line and so belong to a $SL(3,\R)$ subgroup. This is 
related to 
the fact that 1-forms can be dualised to scalars in 3 dimensions.
In fact  for any dimension, and also in type I theories,
1-form potentials in the maximally dualised form belong to the representation 
of highest 
weight equal to  the fundamental dominant weight along  the last root of the 
$SL(10-d)$ line (root labelled $d$,$\, d\le 9$ for type 
I) at least {\it before they can be dualised away}.  

Scalars correspond to the 
adjoint representation of highest weight equal to
the highest root. Their dualisability,
which is a spacetime property, somehow can be seen in the internal symmetry 
Dynkin diagram as the contiguity of the highest weights. 
Any weight is Weyl conjugate to a dominant weight but not necessarily to a 
fundamental one. Conjugacy seems to be a necessary condition for the 
dualisability. For higher forms, the critical dimensions for 
which the degree of  forms can be lowered (3 for vectors to scalars, 5 for 
2-forms to 1-forms etc...) are most visible on the Dynkin diagram when  
dualisability corresponds to the existence of an outer automorphism.

Type II theories provide more examples of this. I have checked them one by one!
The last vertex of the Dynkin diagram to appear (label $d$)
is always associated  to
the highest weight of the vector representation (it is the fundamental weight 
for that vertex), the vertex labeled 9 in 
Figure 1 gives the highest weight of the 2-forms and similarly the vertex
labeled
10 there corresponds to the 3-form. Note that $E_{11-d}$ has outer 
automorphisms visible as symmetries of 
its Dynkin diagram whenever $k$-forms and $(k'=d-2-k)$-forms are in duality, 
this 
symmetry exchanges the locations of the corresponding  (fundamental) weights 
in all 9 cases of duality  for degrees of potential-forms between 0 and 4! 

Let us choose dimension 6 for instance, the group $G$ is 
$SO(5,5)$ and the 2-forms can be selfdual under the symmetry of the diagram 
that 
exchanges the two half-spinor ends, ie the vertices 10 and 6 and leaves 
invariant the vertex 9. These vertices 
correspond to the fundamental weights of the 3-form, dual 1-form and 
self-dual 2-forms 
respectively. The scalars in the adjoint correspond to the lowest root whose 
vertex is adjacent to the vertex 8 related to the potentially dual 4-form as 
we saw at the end of the previous section by a Weyl reflection.
Let us note in passing the strange rule that when the affine diagram is cyclic
($SL(p,\R)^{(1)}\, p\ge 3$) the sum of the degrees of the potential forms 
whose vertex is 
attached to the (affine) extension vertex is its dual degree; for example in 
dimension 8
scalars are dual to 6 forms and 6=4+2, and similarly in dimension 7.

Let us now remark that the outer automorphisms of the internal  Lie algebra 
of  dualities $g$ should be defined as simultaneously dualising the 
world indices. This is quite easy to implement by  similar symmetries of the 
Dynkin diagram of $SL(d)$. It is well known that the  fundamental 
representations of $SL(d)$ are from one end to the other the antisymmetric 
powers of the vector representation in increasing order: the vector, then the 
antisymmetric twice 
contravariant tensor etc... The symmetry under the exchange of the two ends of 
$SL(d)$ is nothing but duality provided we consider field strengths. By this 
we mean that the tensor character of the field strength should define the 
fundamental weight and consequently the  vertex to be dualised. 
Under dimensional reduction and its inverse group disintegration \cite{julia}
the linear group is shared in a $d$-dependent way
 between internal and external spaces. Let us denote 
by a star the scaling  factor $\R$ of symmetry, the general picture for type 
IIA is given in Figure 3.

\centerline{
\begin{tabular}{ccccccccccccccccccc}\\
 $9$& &$8$& & & &$d$& &$w$& &$d-2$& && && &1& &0 \\
 o&---&o&---&o&...&o&   &*&   &o&---&o&...&o&---&o&---&o\\
 &   & &   &$|$&   & &   & &   & &   & &  & &  & &  &\\
 &   & &   &o&   & &   & &   & &   & &   & &  & &  &\\
 &   & &   &$10$& & &   & &   & &   & &   & &  & &  &\\
\end{tabular}}
\bigskip
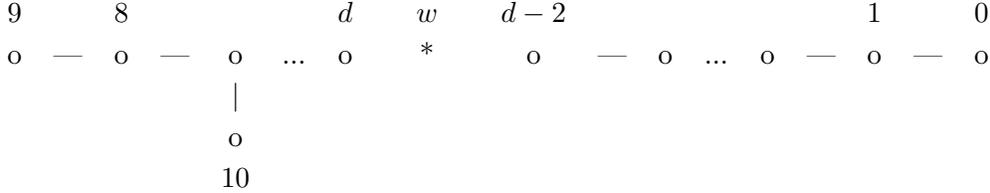
\centerline{Figure 3: type II $F''_{11}$ subgroups of F-symmetry}
\centerline{ The vertex labeled $w$ represents scaling symmetry }
\bigskip

The diagram obtained by adding two 
bonds to Figure 3 to connect a regular $SL(2,\R)$ replacing the $\R$ factor 
to the rest of the diagram contains all   the above diagrams in any
 dimension. Its rank is eleven let us call it $F''_{11}$, it has an analogue 
$F'_{11}$ for type I.
Note that  $SL(12, \R)$ cannot be found in them by naive inspection. However
it could still be included in there, 
like $SL(9,\R)$ inside $ E_8$. But $GL(12,\R)$
is unlikely to be included  and this  could be related to the same property of
F-theory. Note that supersymmetry implies that the twelfth dimension would be 
timelike and hence the subgroup of $SL(12, \R)$ would be $SO(10,2)$ \cite{MG}
so it may be not so surprising that the $SL(12)$ is hidden. 

$F''_{11}$ contains of course  $E_{10}$. $HD_{10}$ lies analogously in 
$F'_{11}$ and M-ology suggests that types I and II
 should be unified to make the full F-group. In fact it is encouraging to 
discover that orbifolding by an involution (a Cartan involution) leads 
in all dimensions from the type II to the type I U-duality. The invariant set
of the involution is given by erasing the vertex 9 of Figure 1 and adding 
vertex 10 of Figure 2. This group theoretical construction should be compared 
to the geometrical orbifolding of \cite{HW}. It is tantalising now to propose 
that the one-loop diagram obtained either by attaching through
 a new single vertex (as the simplest choice) vertices 8 and 10 of 
Figure 2 or vertices 9 and 3 of Figure 1 or 3  should be studied more carefully.
This diagram has rank 12 and a $\Z_2$ symmetry.
 
In summary dualisability seems to require the existence of  (outer)
involutions of the F-duality group of a very special type. The F-group should 
be universal but then split into a  product of spacetime symmetry by an internal
symmetry  factor in a type- and $d$-dependent way.

 Let us now review the possible duality symmetries of a Lagrangian field 
theory in curved space with scalar fields and a N-plet of middle 
rank ($f=d/2$) forms. In \cite{GZ} it was established that in Minkowskian 
signature the set of equations of motion  for the N vector fields 
and sufficiently many scalar fields 
are invariant at most under the symplectic group $SP(2N,\R)$. 
Then the energy-momentum tensor is automatically invariant.
In fact if there is no scalar compensator like in the free Maxwell theory the 
available linear invariance is under $GL(N,\C)$. So
there remains some work to be done to clarify the situation with a few 
scalars. For the free Maxwell case we 
should use the invariance of the energy-momentum tensor as an input instead of 
deriving it as in \cite{GZ} but even then the discussion may be more involved. 
In \cite{CJLP1} the analysis was extended to all dimensions higher than two
and also to Euclidean signature. It is assumed that the half dimension forms
occur at most quadratically in the action and again that there are enough 
scalar fields to be able to reduce the symmetry as in \cite{GZ}.

 The Minkowskian analysis seems to have been 
done a long time ago as we mentioned but the Euclidean case is interesting 
because it illustrates the fact that a change of (non-degenerate) signature 
does not change the duality group; it changes, in general, the coset
however. The non-compactness of the subgroup $H$ that replaces $K$ 
implies some chaos.  
What we have shown is that the case of dimension $d=4k$ resembles that of 4 
dimensions but that of $4k+2$ dimensions however admits maximal duality
group of the type $O(N,N)$ with compact subgroup (Lorentzian case) $O(N)^2$
resp. subgroup $H=O(N,\C)$. For completeness let us mention that in $4k$
dimensions $K=U(N)$ is similarly replaced in the Euclidean signature by
$H=GL(N,\R)$. 

This analysis was motivated in part by my desire to 
clarify the difference between two-dimensional quantisation of single charges
by requiring locality
(symmetric quadratic form) and the four-dimensional Dirac-Schwinger-Zwanziger 
antisymmetric quantisation condition. As we have seen this alternation between 
symmetry and antisymmetry of the quadratic invariant $\omega$ for every other 
dimension can be checked by 
inspection and is rather robust as it does not depend on the signature of
spacetime. The supersymmetric origin of the supergravity bosonic actions is not
relevant here but let us recall that dualities are
 often coupled to chiral rotations
in such models. Now it has been noticed repeatedly that the first nontrivial 
example of what we would like to call a quantisation dimension namely 6 is 
closely related to quaternions and hyperKaehler structures \cite{KT}.
But $k$ was arbitrary in the previous discussion so we should not restrict 
ourselves to the supersymmetry domain.

One conclusion at this stage is that, as one had maybe anticipated, the 
symmetry must be defined by a very basic counting, namely parity of
 the half-dimension $f:=d/2$; this number precisely determines 
the symmetry or antisymmetry of the intersection form of the corresponding 
cohomology. An important shift of one occurs though as $f=1$ in the symmetric 
case here, we shall elaborate on it later, it is related to the time extension. 

Let us finish this section by giving some technical advice for the handling of 
non-manifestly duality invariant actions. It was inspired by \cite{BB} and 
consists in doubling as usual the set of forms of degree $f$ as we explained 
above and 
in keeping along the TS constraint which is also invariant.  
This leads to nicer couplings to scalar fields
\cite{CJLP1} for instance and to a derivation of the Noether current of duality which had to be guessed in \cite{GZ}.

\bigskip
\section{Complementarity, dyons, TS and F-duality and conclusion}

Let us now elaborate on the relationship between charge quantisation conditions
for dyons in 4 dimensions and for chiral fermions in 2d. I noticed  in March 
1996 
that the famous Skyrme minus sign, namely the fermionic character of his
vertex operator had to be understood in units where $h=2\pi$ as
$$ e^{2i\pi (eg'+e'g)} = -1$$ 
This should have been (and maybe has been) analysed long ago. The analogy 
between chirality of spinors and helicity of vectors is a Lorentz fact. As we
mentioned above supersymmetric dualities act simultaneously on the fermionic
fields by chiral transformations. As it appears from \cite{BMT} chiral fermions
are obtained by a  vertex construction if their charge obeys the quantisation 
condition: $e=g=\sqrt{s} /2$, where $s$ is an integer (a squarefree integer
for the algebra of observables to be maximal). The model is the chiral $U(1)$ 
current algebra.  The idea is that a nonlocal 
expression can only satisfy causality, or fermionic causality, if it obeys a
quantisation condition \` a la Dirac. In the monopole case one wants the
Dirac string to be invisible, in the fermionisation problem one wants the 
Mandelstam-Skyrme string to be almost invisible, giving only a minus sign but 
actually an unavoidable one.  
In the Sine-Gordon case the quantisation is the celebrated $\beta^2 =4 \pi$ 
relation. It could be rewritten, compare \cite{SW}, as the relation 
$eg'=\frac{1}{4}$ but the symmetry of the quadratic form was overlooked there 
and generalisations attempted in 4 dimensions.  We now know much more about 
affine Kac-Moody algebras (then 2 years old in Mathematics and to be born in 
Physics). The finiteness of the Schwinger term in 1 (null) dimension or in 
1+1 dimensions is the possibility of central extension of Loop groups. 
A lot of effort has been devoted to find higher dimensional generalisations
but divergences have spoiled the game. Yet non-relativistic versions of 2+1
or 3+1 dimensional
fermionisation have been proposed since 1975 in the presence of gauge fields.
The role of the Dirac quantisation condition appeared immediately through the
formula for angular momentum, which gives straightforwardly the minus
sign of the Schwinger-Zwanziger condition. It is very natural to compare this 
to the Bohr-Rosenfeld uncertainty relation for electromagnetic fields. This is 
in 4 dimensions but free field theory and hence (bad)-divergence free. 

Interestingly enough the quantisation condition for extended objects (p-branes)
with $p\ge 2$ were given in 1985-86 without adressing the question of the sign.
It was thus very exciting to check that in 6d supergravities ($N_4=8 $ and 
$N_4=6$ \cite{julia}) the 
duality symmetry was not symplectic but pseudo-Riemannian.
Since then a few papers have appeared to fill the gap mentioned above by direct
arguments and to verify the sign difference between $4k$ and $4k+2$ 
dimensions for the generalised Dirac-Schwinger-Zwanziger condition \cite{GANG}.
One may remark that these papers exclude the fermionic case.
Now dimensional reduction techniques are available to exploit a desired
fermionisation that could be done in six dimensions without any internal 
symmetry, so as to do it in lower 
dimensions with an automatic twist. The simplest candidate is a two-form 
(self-dual if one prefers) in 6d and in fact such a theory has become quite 
fashionable in the meantime starting with \cite{VW}. We 
can check that the fibration of the 6d Minkowskian space on a torus over
either Euclidean or Minkowskian base gives in both cases the $SL(2,\Z)$ modular
group as duality group on the base for Maxwell theory. 
It was clear already from the disintegration magic of \cite{julia} that the 
rather fundamental nature of the so-called ADE structure seemed to play a role 
in supergravity theory. Since then it appeared in CFT, in the applications of 
singularity theory (where it was already a dominant character) to string theory,
etc... Its home, 
intersection theory seems to be  the most important tool for 
fermionisation and charge quantisation problems \cite{PRJ}.

\bigskip

The second project I have been interested in that relates to dualities is the 
full analysis of the 1975 observation that classical field theory is obtained
by letting Planck's constant tend to zero but hiding some of them in rebaptised 
parameters like E=e/h and M=m/h. The upper case parameters are relevant to the 
wave world in which one does not count quanta. Note that e/m=E/M, so the 
anniversary of the electron as a particle should have been next year instead of 
last year. Now this relates to dualities because in this (electric) classical 
field theory limit one can study magnetic particles and their solitonic 
``quantised" charge. Clearly the dual theory, ignoring its strong coupling 
for the moment, has magnetic waves and electric particles at the new classical 
level. One message is that the  words ``classical limit" are so ambiguous that
they confuse everybody. Let us mention in passing that non-relativistic limits
are equally ambiguous as anybody at ease with Unit Systems can testify.
How do classical people measure Planck's constant? They in fact combine two 
complementary limits to measure for instance two vertices of
the [e, E or g, $\alpha$] triangle.
The validity of a particular classical limit is to be decided case by case and
one may be better than another. Is it true that the collection of all classical 
limits contains all  the information about the quantum theory? This is  
postulated in the usual interpretation of Quantum Mechanics, but 
can we make this precise and restrict the set of limits that is 
necessary?
For instance non-relativistically invariant classical limits are sometimes 
necessary for instance to obtain particles rather than quanta in order to 
avoid the Klein paradox. Similarly non-relativistically invariant limits of 
supergravity theories might be useful for some perturbative computations.  

To conclude with less ambitious goals let me sketch some results of \cite{CJLP2}
in order to motivate the reader to dig deeper into the present lines. It turns 
out that indeed the TS idea can be implemented for all differential forms of the
type II supergravity theories in all dimensions between 3 and 11. The structure is particularly simple of course in 11 dimensions but the equations can be 
rewritten in a dimension independent way. It applies to sigma models both gauge 
fixed and not gauge fixed. The restriction to toroidal compactifications remains
and  should be relaxed in the future. Our present reduction of type 
II to type I suggests that this might be doable. Bosonic strings and heterotic
strings should be reconciled with the present group theoretical approach that
smells of integrable systems, in a loose sense as chaos generally follows 
non-compact groups. 

\bigskip
{\bf Acknowledgements} I would like to thank P. Windey for inviting me to
present some of these results at the NATO Advanced Study Institute on 
``Strings, Branes, and Dualities". 
Parts 3 and 4 benefited from many discussions with 
E. Cremmer, part 5 from encouragements from and discussions with 
D. Buchholz, K. Fredenhagen and  J. Roberts in 1995 and  1996 resp.
C. Callan in 1975 and H. B. Nielsen in 1993 and 1994. 
I am also grateful for useful  
comments to C. Bachas, Y. Benoist, D. Bernard, M. Duflo, M. Henneaux,
C. Hull, H. Nicolai, D. Olive, M. Trigiante. This work was
supported in part by EEC under TMR contracts ERBFMRX-CT96-0045 and -0012 
and by the A. von Humboldt foundation.

\end{document}